\newcommand{\xfrac}[2]{{#1}/{#2}}

\newcommand{\blk}{\color{black}}

\newcommand{\beq}{\begin{equation}}
\newcommand{\eeq}{\end{equation}}
\newcommand{\bqa}{\begin{eqnarray}}
\newcommand{\eqa}{\end{eqnarray}}
\newcommand{\nn}{\nonumber}

\newcommand{\smallfrac}[2]{\mbox{$\frac{#1}{#2}$}}

\newcommand{\sch}{Schr\"odinger}

\newcommand{\half}{\smallfrac{1}{2}}

\newcommand{\cu}[1]{\left\{ {#1} \right\}}

\documentclass[journal,article,submit,moreauthors,pdftex,10pt,a4paper]{mdpi}

\usepackage{placeins}
\usepackage{amsmath}

\firstpage{1} 
\articlenumber{x}
\doinum{10.3390/------}
\pubvolume{xx}
\pubyear{2016}
\copyrightyear{2016}

 \theoremstyle{mdpi}
 \newcounter{thm}
 \newcounter{ex}
 \newcounter{re}
\Title{Nonlocality in Bell's theorem, in Bohm's theory, and in Many Interacting Worlds theorising}

\Author{Mojtaba Ghadimi$^{1}$, Michael J. W. Hall$^{1}$, 
and Howard M. Wiseman$^{1,*}$}

\AuthorNames{Mojtaba Ghadimi, and Michael J. W. Hall, 
and Howard M. Wiseman}

\address{%
$^{1}$ \quad Centre for Quantum Dynamics, Griffith University,
Brisbane, Queensland 4111 Australia. 
}
\corres{Correspondence: h.wiseman@griffith.edu.au} 

\abstract{``Locality'' is a fraught word, even within the restricted context of Bell's theorem. As one of us has argued elsewhere, that is partly because Bell himself used the word with different meanings at different stages in his career. The original, weaker, meaning for locality was in his 1964 theorem: that the choice of setting by one party could never affect the outcome of a measurement performed by a distant second party. The epitome of a quantum theory violating this weak notion of locality (and hence exhibiting a strong form of nonlocality) is Bohmian mechanics. Recently, a new approach to quantum mechanics, inspired by Bohmian mechanics, has been proposed: Many Interacting Worlds. While it is conceptually clear how the interaction between worlds can enable this strong nonlocality, technical problems in the theory have thus far prevented a proof by simulation. Here we report significant progress in tackling one of the most basic difficulties that needs to be overcome: correctly modelling wavefunctions with nodes.}


\begin{document}

\section{Introduction}\label{sec:in}

This paper is based loosely on the talk given by its final author at the \blk 2017 Symposium on Emergent Quantum Mechanics (EmQM17). \blk
That~talk took as its principal inspiration the following questions (from the list provided by the organisers to give focus to the meeting): 
\begin{itemize}[leftmargin=*,labelsep=5mm]
\item Is the universe local or nonlocal? \vspace{1ex} 
\item What is the future of scientific explanation? Is scientific metaphysics, e.g., the notions of reality, causality, or physical influence, obsolete in mathematical accounts of the quantum world?
\end{itemize}

However, in view of the conference also being the David Bohm's Centennial Symposium, the speaker also briefly discussed some work relating to this question: 

\begin{itemize}[leftmargin=*,labelsep=5mm]
\item What is David Bohm's legacy for the future of quantum physics? 
\end{itemize}

This paper also concerns these two broad topics, but in the opposite ratio to that in the delivered talk. Section \ref{sec2} of this paper addresses just the question ``Is the universe local or nonlocal?'' by examining the historical meaning of the term ``local''. (The second question listed above, which was addressed in  the talk, and in References~\cite{Wis14a,Wis14b,WisCav17}, is not addressed here). That section expresses the views of the invited contributor, not necessarily those of all the authors.  Sections \ref{sec3}--\ref{sec6} concern Many Interacting Worlds (MIW), an approach to quantum mechanics inspired by Bohm's original hidden variables theory~\cite{Boh52_1,Boh52_2}, and which thus we hope will be part of the answer to the last question above. 

Having introduced the MIW approach in Section \ref{sec3}, we turn in Section \ref{sec4} to 
how it can give rise to nonlocality; that is, how it can offer a positive answer 
to this (here abridged) focus question: 
\begin{itemize}[leftmargin=*,labelsep=5mm]
\item Are nonlocal connections---e.g., ``action-at-a-distance''---fundamental elements in a radically new conception of reality?
\end{itemize}

However, as Section \ref{sec4} notes, there are still major difficulties faced in demonstrating these connections in MIW simulations. Section \ref{sec5} introduces a new MIW technique---a higher-order interaction potential---to address one of these difficulties, namely, wavefunction nodes. Section~\ref{sec6} applies this idea to progress the MIW simulation of stable excited state nodes of a quantum particle in one dimension. {An alternative method to address the node problem is presented and simulated in Appendix} \ref{Equivariance}. Section \ref{sec7} concludes with a discussion of open challenges for {MIW theorising}. 

\section{Is the Universe Local or Nonlocal?} \label{sec2}

The answer to this EmQM17 focus question hinges, of course, on what one means by the term ``local'' (assuming that ``nonlocal'' is simply its complement). 
Perhaps surprisingly, it seems~\cite{Wis14b,WisRie15} that the word was not used in the context of interpreting {EPR} quantum correlations prior to the 1964 paper of Bell~\cite{Bel64}. 
In that paper, Bell proved his 1964 Bell's theorem; to quote~\cite{Bel64}:  
\begin{quote}
In a theory in which parameters are added to quantum mechanics to determine the results of individual measurements, without changing the statistical predictions, there must be a mechanism whereby the setting of one measuring device can influence the reading of another instrument, however remote. 
\end{quote} 

In other words, some quantum phenomena are incompatible with the joint assumption of predetermination (or causality; Bell used both terms) and 
locality (or separability; Bell used both terms). In the above quote, it is the negation of locality that is characterised, in a way consistent with Bell's earlier definition of locality; to quote~\cite{Bel64}:
\begin{quote} 
It is the requirement of locality, or more precisely that the result of a measurement on one system be unaffected by operations on a distant system with which it has interacted in the past, that creates the essential difficulty.
\end{quote}

Thus, Bell intended to be (somewhat) precise about what he meant by locality. Unfortunately 
he did not give a general mathematical definition, nor did he define terms like ``unaffected'' or (in the first quote and elsewhere) ``influence''.  

In the theorem he proves that the role ``locality'' plays is the following. The assumption of predetermination means that an arbitrarily long time before any measurements are performed, there existed in the world a collection of hidden variables, $\lambda$, that, together with the future measurement settings, determines the future outcomes. Adding the ``vital assumption [of locality]'' 
 implies that the outcome $A$ of one party (Alice, say) cannot depend on the settings $b$ of a distant party (Bob, say), but only by her own setting $a$. That is, in a theory $\theta$ with local predetermination of outcomes, there exists a function, $A_\theta(a,\lambda)$ such that $A=A_\theta(a,\lambda)$, and likewise for Bob. From this, Bell was able to derive his famous theorem, that there exist sets of measurement on entangled quantum systems whose results cannot be explained by any such model. ({{Note that Bell made an implicit assumption in 1964, related to freedom of choice, that $P(\lambda|a,b) = P(\lambda)$. Here we follow Bell, as such an assumption is certainly necessary}~\cite{MJW_measure}.})

In the above formulation, ``locality'' has a precise meaning only in the context that predetermination has already been assumed. This is fine for the purpose to which Bell puts it. If one wanted to broaden the definition so that it applied independently of the assumption of predetermination then, it has been argued~\cite{Jar84,How85,Wis14b,WisRie15}, the natural reading of Bell's verbal definition would be as follows. An arbitrary theory contains initial variables $\lambda$, which may or may not be hidden, and which may or may not be sufficient to determine all outcomes, and is described (in the limited context we are considering) by theoretical probabilities $P_\theta(A,B|a, b, \lambda)$. 
The theory is local if there exists a function 
$P_\theta(A|a, \lambda)$ such that 
\beq \label{Deflocal} 
 \forall\, b,\ \blk P_\theta(A|a, b, \lambda) = P_\theta(A|a, \lambda).
\eeq
and likewise for Bob. 

Others have argued~\cite{DGZ92,Mau94,Nor06,Nor15}, to the contrary, that this definition does not fit with the other use Bell makes of the concept of locality in his 1964 paper, which is the first paragraph of his Section~II. There~Bell implies that, according to the EPR argument~\cite{EPR35}, locality plus perfect correlations of outcomes implies predetermination. This shows, if nothing else, that Bell did think, in 1964, that locality was an assumption that had meaning prior to the assumption of predetermination. However,~the obvious meaning, Equation \eqref{Deflocal}, does not work in the EPR argument. Even one of Bell's most ardent admirers~\cite{Nor06} was forced to admit this~\cite{Nor15}:
\begin{quote}
It is simply not clear how to translate Bell's
words here (about locality) into a sharp mathematical statement in terms of which the EPR argument might be rigorously rehearsed. \dots {[I]t} must be admitted that Bell's recapitulation of 
the EPR argument in this paragraph leaves something to be desired.
\end{quote}

Regardless, is there really a problem here? Bell does not say that he believes that locality plus perfect correlations implies predetermination. He merely says, at the beginning of that paragraph, ``\dots the EPR {\em argument} is as follows''  (my emphasis). In the preceding (opening) paragraph he is even weaker: ``The paradox of Einstein, Podolsky and Rosen was {\em advanced as an argument} \dots'' (my emphasis again). In a follow up paper in 1971~\cite{Bel71}, Bell is weaker still. He gives three motivations for the assumption of predetermination (in a section entitled, unambiguously, ``Motivations''). An EPR-style argument is the third motivation; he does not claim any logical deduction from locality but merely appeals to the intuition of the reader for the reasonableness of predetermination by hidden variables (see Reference~\cite{WisRie15} for details). The 1969 paper of CHSH~\cite{CHSH69} seems actively skeptical of the EPR argument, saying it ``{\em led them to infer} that quantum mechanics is not a complete theory'' (my emphasis). 
 It is thus clear that, at this time, the EPR argument was far from being regarded as a rigorous proof for the necessity of hidden variables. (For a discussion of  the extent to which the EPR argument is such a proof, see Reference~\cite{Wis13}.) 
 
 The situation with regard to the EPR argument changed rapidly, at least in Bell's mind, after he formulated, in 1976~\cite{Bel76}, a concept that does allow one to infer predetermination from perfect correlations. This was the concept of ``local causality'', stated most succinctly in a later paper~\cite{Bel90b} 
 \begin{quote}
 A consequence \dots of ``local causality'' {[is]} the outcomes  {[in the two labs]} having no dependence on one another nor on the settings of the remote  {[measurement]}, but only on the local  {[measurement settings]} and on the past causes. 
\end{quote}

 In the situation considered above, a theory is ``locally causal'' only if there exists a function  
$P_\theta(A|a, \lambda)$ such that 
\beq \label{DefLC} 
 \forall\, b,B,\ \blk P_\theta(A|a, B, b, \lambda) = P_\theta(A|a, \lambda),
\eeq   
and likewise for Bob. 
Note the appearance of $B$ as a conditional variable on the \blk left hand side, \blk 
which distinguishes this concept from ``locality'' 
 in Equation \eqref{Deflocal}. Moreover, this assumption obviates the need to consider determinism at all---it leads directly to the Bell inequalities that quantum mechanics violates. Thus, Bell gave  what I have called~\cite{Wis14b} Bell's second Bell's theorem, in 1976~\cite{Bel76}: 
\begin{quote}
Quantum mechanics\dots gives certain correlations which \dots cannot be {[reproduced by]} a locally causal theory. 
\end{quote}

Bell clearly (and, I think~\cite{Wis14b,WisCav17}, rightly) thought local causality to be a more natural concept than locality as per Equation \eqref{Deflocal}, as he never used the latter concept again. Regrettably, however, he did not abandon the word ``locality''. Rather, beginning even in 1976~\cite{Bel76}, he sometimes used ``local'' as short-hand for ``locally causal'', and, a few years later, was apparently convinced that local causality was the concept that he (and EPR) had always used~\cite{Bel81} (for details, see~\cite{Wis14b}). However, at least in his final word on the subject \cite{Bel90b}, Bell showed his preference 
unequivocally for the terminology ``local causality'' over ``locality''. 

Thus we may return to the primary question posed above---{\em Is the universe local or nonlocal?} 
  If~by ``locality'' one means ``locally causal'', the concept Bell promoted for most of his career in quantum foundations~\cite{BellCollection}, then the answer (barring more exotic possibilities such as 
``superdeterminism''~\cite{Bel90b}, retrocausality~\cite{price08}, and the subjectivity  of macroreality~\cite{Deutsch99}) is that the universe is nonlocal; it violates local causality.  To avoid confusion, we might agree to say that the universe is Bell-nonlocal~\cite{Wis14b}. If,~on the other hand, one adopts the definition of ``local''  indicated by Bell's 1964 paper and commonly used in text books~\cite{NieChu00,SchWes10}, then the answer is that we cannot say whether the universe is local or nonlocal. Operational quantum mechanics satisfies this weaker sense of locality, simply because it does not feature signalling faster than light, and denies the need for giving any account for quantum correlations beyond an operational one. We can only say that the universe is nonlocal, in this strict sense, if we make some other assumptions about its nature, such as determinism.

 \section{David Bohm's Legacy: Permission to Theorise Radically New Conceptions of Reality} 
 \label{sec3}
 
 The most famous quantum theory which does make the assumption of determinism is of course David Bohm's~\cite{Boh52_1,Boh52_2}. Indeed, Bohm's theory was an inspiration to Bell who summarised the result of his 1964 paper in the introduction-cum-abstract as 
 \begin{quote}
{[A]} hidden variable interpretation of elementary quantum 
theory~\cite{Boh52_1,Boh52_2} has been explicitly constructed. That particular interpretation has \dots a grossly nonlocal
structure. This is characteristic, according to the result to be proved here, of any such theory which
reproduces exactly the quantum mechanical predictions.
 \end{quote}
 
Why Bell considered Bohm's interpretation to be ``grossly nonlocal'', rather than nonlocal {\em simpliciter},  
 is unclear. Perhaps it was because the theory is nonlocal even in situations where there is an obvious local hidden variable theory, as in the EPR-correlations~\cite{EPR35}, or the EPR-Bohm correlations~\cite{Boh51}. 

Unlike operational quantum mechanics, Bohm's theory is a precise and universal physical theory. Restricting to the case of interacting nonrelativistic scalar particles for simplicity of discussion, it takes the universe to be described by a universal wavefunction $\Psi({\bf q})$, obeying \sch's equation, where ${\bf q}$ is the vectorised list of the coordinates of all the particles. However, it also postulates a single point in configuration space, ${\bf x}$, which encodes the {\em real} positions of all these particles. This ``marvellous point''~\cite{Albert15} or ``world-particle''~\cite{MIW14} has a deterministic equation of motion $\dot{\bf x} = {\bf v}_\Psi({\bf x})$ guided  vicinally  by $\Psi({\bf q})$. ({{Note that ``vicinal'' is a synonym of ``local'' in the latter's quotidian sense, introduced here to avoid any possible confusion with ``local'' in the technical sense defined in Section}~\ref{sec2}.}) 
Here, ``vicinal guiding'' means that the world-particle's \blk velocity ${\bf v}_\Psi({\bf x})$ depends on $\Psi({\bf q})$ and finitely many derivatives, evaluated at ${\bf q} = {\bf x}$. However, vicinal \blk in configuration space is not  vicinal \blk in 3D space---the  positions of Bohmian \blk particles in one region of 3D space 
can affect the motion of an arbitrarily distant particle if entanglement is present.  Since it is the position of Bohmian particles that encodes what an experimenter decides to measure, this gives rise to the `gross' nonlocality of Bohmian mechanics which Bell noted in 1964. 

Bohm's original proposal~\cite{Boh52_1,Boh52_2} actually used a second-order dynamical equation $\ddot{\bf x} = {\bf a}_{\rm  vicinal \blk}({\bf x}) + 
{\bf a}_{\Psi}({\bf x})$. The 3D-vicinal \blk acceleration ${\bf a}_{\rm  vicinal \blk}({\bf x})$ is given by Newton's laws,  involving inter-particle potentials which drop off with 3D-distance. The nonlocal effects in Bohm's theory arise from a separate, 3D-nonvicinal, quantum acceleration \blk ${\bf a}_{\Psi}({\bf x})$. 
Bohm's publication of an explicitly nonlocal theory seems to have made it acceptable for other physicists to publish realist approaches to quantum mechanics in direct opposition to the Copenhagen interpretation. This included: de Broglie in 1956~\cite{deB56}, reviving his unpublished idea from 30 years earlier which prefigured much of Bohm's work and used the first-order dynamics described earlier; and Everett in 1957~\cite{Eve57}, introducing the relative state interpretation, more popularly known as the many worlds interpretation. 

Taking inspiration from both Bohm and Everett, two of us plus Deckert introduced in 2014~\cite{MIW14} what we called the Many Interacting Worlds (MIW) approach to quantum mechanics. We suggested that it might be possible to reproduce quantum phenomena without a universal wavefunction $\Psi({\bf q})$  (except to define initial conditions). In its place we postulated an enormous, but countable, ensemble ${\bf X} = \cu{{\bf x}_j:j}$ of points ${\bf x}_j$ in configuration space (similar ideas have been proposed earlier by a number of authors~\cite{Holland05,Poirier10,ParPoi12,SchPoi12}, but they considered a continuum of worlds, which, in our view, leads to some of the same conceptual issues that Everett's interpretation faces---see also~\cite{sebens}). Each~point is a world-particle, just as Bohmian mechanics postulates,  and the dynamics is intended to reproduce a deterministic Bohmian trajectory for each world-particle. However, the trajectory of a world-particle is guided not by a wavefunction, but by the locations of nearby world-particles \blk in configuration space: \mbox{$\ddot{\bf x}_j = {\bf a}_{\rm  vicinal \blk}({\bf x}) + {\bf a}_{\rm MIW}({\bf N}_j)$}, where {${\bf N}_j \subset {\bf X}$} 
 is the set of world-particles in the  vicinity \blk (somehow defined) of world-particle ${\bf x}_j$, and ${\bf a}_{\rm MIW}$ is some fixed ($j$-independent) function describing the interaction of many worlds.  

In the MIW approach, probabilities arise only because observers are
ignorant of which world ${\bf x}_j$ they actually occupy, and so assign an equal weighting to all worlds compatible with the  macroscopic state of affairs they perceive, in accordance with Laplace's principle. In a typical experiment, where the outcome is indeterminate in operational quantum mechanics, the final configurations of the worlds in the MIW approach can be grouped into different subsets, {${\bf X}_o \subset {\bf X}$} 
 of world-particles,  still extremely large in number, based on shared macroscopic properties corresponding to the different possible outcomes $o$. In Everett's approaches, these groups correspond to branches of the universal wavefunction, and one has to argue that the square modulus of the coefficient of each branch somehow manifests as the correct probability for the experimenter. In the MIW approach, the operational quantum probabilities will be equal to the number of worlds in each group divided by the number of worlds at the start.

\section{Nonlocality in the Many Interacting Worlds Approach}\label{sec4}
\vspace{-6pt}
\subsection{ General Considerations}

Just as Bohmian dynamics is  vicinal \blk in configuration space but  gives rise to nonlocality \blk in 3D space, so too is the type of MIW dynamics just described.  Indeed, it must be nonlocal because, like Bohmian mechanics, it is deterministic. However, we can be more specific about how this nonlocality, in Bell's 1964 sense, arises in the MIW approach. \blk

Consider a branch $\Psi_E$ (in the Everettian sense) of the universal wavefunction $\Psi$.  In \blk the MIW approach, such a branch will correspond to a subset ${\bf X}_E \subset {\bf X}$ of world-particles,  still extremely large in number, that are close together (on a  macroscopic scale) in configuration space and mostly far (on a macroscopic scale) from  the rest of ${\bf X}$. \blk 
Thus, to a good approximation, this set of world-particles will evolve autonomously, on the time scale of interest. 

 Now consider a particular world-particle, ${\bf x}_j \in {\bf X}_E$, and a particular part of the vector ${\bf x}_j$, comprising a lower-dimensional vector ${\bf b}_j$, that contains the variables that \blk encode the macroscopic fact of the decision by an experimenter, Bob, about what experiment to perform. That is, the decision is the same in every world under consideration here. However, there is no reason in the theory for ${\bf b}_j$ to be correlated with the other variables in ${\bf x}_j$, until the experiment is actually performed. Once it is performed, the value of ${\bf b}_j$ will have a direct (second-order in time) effect on the  vicinal \blk (in the 3D sense) variables in ${\bf x}_j$, say ${\bf B}_j$. Then, in the MIW approach, the change in ${\bf B}_j$ will cause a change (again second-order in time) in ${\bf x}_k \in {\bf N}_j$  (note that ${\bf N}_j$ is only a tiny subset of ${\bf X}_E$). 
Now say that in $\Psi_E$, there is entanglement between the system corresponding to ${\bf B}_j$, and a distant system corresponding to some other observables ${\bf A}_j$. 
Then the change in ${\bf x}_k$ will be a change not just \blk in ${\bf B}_k$ but also in ${\bf A}_k$. This~is because the MIW acceleration function ${\bf a}_{\rm MIW}$ pays no heed to  vicinality \blk in the 3D sense---it cares only about the  fact that the worlds ${\bf x}_k\in {\bf N}_j$ are in the vicinity of the world ${\bf x}_j$ \blk in configuration space. Finally,~the same interaction will cause a (second-order in time) change in ${\bf A}_j$, the distant observables in the original world $j$, since for at least some of the ${\bf x}_k \in {\bf N}_j$ we will have ${\bf x}_j \in {\bf N}_k$.  

Thus it is clear in principle how the MIW approach can give rise to nonlocality in Bell's 1964 sense, with Bob's choice ${\bf b}_j$ causing a change in the  variables ${\bf A}_j$, variables which correspond to the outcome observed by Alice in her distant, perhaps even space-like separated, lab, \blk all in the same world $j$.  It~might seem that Bob could use this to signal to Alice faster than light. However, that is not so, for essentially the same reason that signal-locality is respected in Bohmian mechanics. Alice and Bob are ignorant of which world $j$ they inhabit. They know only the macroscopic configuration ${\bf X}_E$ in which their world must lie. A successful MIW theory would thus exhibit nonlocality at the hidden level of an individual world, but not at the observable level, averaging over all the worlds in ${\bf X}_E$. \blk

\subsection{ Simulations}

Can we explicitly simulate this in our MIW approach, showing, for example, the violation of a Bell inequality, or even just the EPR paradox,  without signalling? \blk Here we run up against the limitations in the development of our approach. 

In our 2014 paper~\cite{MIW14}, we introduced a toy MIW theory to describe a   very simple universe,  comprising a single particle in 1D. 
We showed analytically that, in the limit of a large number of worlds, it gave the correct ground state distribution for a harmonic oscillator, and the correct description for the first and second moments of free particle evolution. We showed numerically that it could 
reproduce, at least qualitatively, the double-slit interference phenomenon. Furthermore, we argued that it can plausibly reproduce other generic quantum phenomena such as barrier tunnelling and~reflection. 

To demonstrate nonlocal correlations between measurements on distant quantum systems, we obviously require more than one particle. One might think one would require particles to model the measuring apparatuses as well as having at least two entangled particles. However, since the position of a particle is reified in the MIW approach,  similarly to Bohm's theory~\cite{Boh52_1,Boh52_2}, a particle can always represent its own measured position   (this differs from its momentum, as any measurement of momentum will rely on the action of the quantum potential in Bohm's theory, or the interworld interaction potential in our approach).  Moreover, if we consider a particle on a spring, with an externally controllable spring constant $k$, then either its initial position or its initial momentum can be encoded in position at a nominated time. This is because such evolution turns momentum information into position information at time $T/4 = (\pi/2)\sqrt{m/k}$, and back into position information at time $T/2$.
 ({{Alternatively, the same can be achieved even by free evolution with an externally controlled mass $m$, an assumption which does not violate any of the MIW framework. By reducing the mass, the time taken for the momentum to be 
encoded into the position can be made arbitrarily small; by increasing the mass, the position remains as it was initially to an arbitrarily good approximation. Thus, either 
position or momentum could be measured at a nominated time}.})  In this way, nonlocal correlations can be explored with a world of just two 1D particles. This scenario could be used not only to demonstrate the EPR paradox, but also Bell nonlocality, as proven by Bell himself in 1986~\cite{Bel86}. 

Recently, we have generalised MIW for a 2D particle, which is equivalent to two 1D particles, with some numerical success~\cite{HHWD17}. However, that study was restricted to finding ground states---stationary states with no nodes in the wavefunction. The same paper did consider finding the first excited states for a 1D particle, but only for symmetric potentials for which the node at $x=0$ can be put in by hand. This  restriction was necessary because, it is now clear, the toy MIW model of Reference~\cite{MIW14} cannot reproduce stationary excited states at all. More generally, it cannot be expected to quantitatively reproduce dynamical quantum evolution in which nodes appear and disappear.  In the proposal of Bell~\cite{Bel86}, to violate a Bell inequality by free evolution of two entangled particles, the required initial wavefunction has two nodes and can be expected to develop more during evolution. Thus,  to study nonlocality in the MIW approach we certainly need to go beyond the toy model of Reference~\cite{MIW14}. Progress in this direction is the topic of the next two sections.

\section{MIW Beyond the Toy Model} \label{sec5}

The MIW toy model, for a nonrelativistic 1D quantum particle of mass $m$ moving in a potential $V(q)$, has $N$ worlds each containing a 1D particle. Denoting the positions and momenta of these world-particles by $X=(x_1,x_2,\dots ,x_N)$ and $P=(p_1,p_2,\dots, p_N)$, their motion is described in the toy model by a Hamiltonian of the form~\cite{MIW14}
\beq \label{hxp}
H(X,P) = \sum_{n=1}^N \left(\frac{(p_n)^2}{2m} + V(x_n)\right) + U_{\rm toy}(X) .
\eeq

The first term is just the sum of $N$ classical Hamiltonians, one for each world. It is the second term, a potential energy responsible for interactions between the worlds, which accounts for quantum~phenomena:
\beq \label{utoy}
 U_{\rm toy}(X) = \frac{\hbar^2}{8m} \sum_{n=1}^N \left(\frac{1}{x_{n+1}-x_{n}} -  \frac{1}{x_{n}-x_{n-1}}\right)^2 .
 \eeq
 
For evaluation purposes we formally define $x_0:=-\infty$ and $x_{N+1}=\infty$, and the ordering $x_1<x_2<\dots<x_N$ has been assumed (this ordering is preserved under time-evolution, due to the repulsive nature of the potential). The motion of each world-particle is determined by the usual Hamiltonian equations of motion, i.e., 
\beq \label{hameq}
\dot x_n = \frac{\partial H}{\partial p_n} =m^{-1} p_n,\qquad\qquad \dot p_n = - \frac{\partial H}{\partial x_n} = -V'(x_n)-\frac{\partial U_{\rm toy}}{\partial x_n} .
\eeq

Note that the interworld potential $U_{\rm toy}(X)$ vanishes in the classical limit $\hbar=0$, and also for the case $N=1$. In either of these cases, each world evolves independently, according to Newton's laws. More generally, however, for $\hbar\neq 0$ and $N>1$ the interworld potential $U_{\rm toy}(X)$ leads to forces on each world that act to reproduce quantum phenomena such as Ehrenfest's theorem, spreading of wave packets, tunneling through a barrier, and interference effects~\cite{MIW14}.  For the case of a 1D oscillator, $V(q)=\half m\omega^2 q^2$, it has further been shown,  in the limit $N\rightarrow\infty$, that the average energy per world of the MIW ground state converges to the quantum groundstate energy $\half\hbar\omega$~\cite{MIW14}, and that the corresponding stationary distribution of worlds samples the usual quantum Gaussian probability distribution~\cite{MIW14,McKeague}.
 
 The form of $U_{\rm toy}(X)$ in Equation~(\ref{utoy}) above is a sum of three-body terms, leading to a force on the $n$-th world that depends on the positions of the two neighbouring worlds on either side. However,~while this form is sufficient to reproduce the quantum phenomena noted above, we have found that it is too simple to model the behaviour of quantum wave functions with nodes. Hence, as noted in the previous section, we must turn to more complex forms of the interworld potential,  involving interaction between greater numbers of neighbouring worlds.
 
 Fortunately, there is a great deal of freedom in choosing this potential in the MIW approach. It is possible this freedom could be curtailed via suitable physically-motivated axioms (such as, for example, requiring Ehrenfest's theorem to hold).  Here, however, we take a nonaxiomatic approach, to show how the interworld potential of the toy model can be straightforwardly generalised to allow direct interactions between an arbitrary number of worlds, in a manner corresponding to greater accuracy in approximation of the Bohmian acceleration.  This greater accuracy supports a corresponding expectation of being able to successfully model wave function nodes.
 
 In particular, it was shown in section~II.D of Reference~\cite{MIW14} that a suitable general form for the interworld potential in the 1D case is obtained by replacing $U_{\rm toy}(X)$ in the Hamiltonian (\ref{hxp}) by
 \beq \label{ux}
 U(X) = \frac{\hbar^2}{8m} \sum_{n=1}^N  \left( \frac{P'_n}{P_n}\right)^2 ,
 \eeq
 where $P_n$ and $P'_n$ are approximations to a probability density $P(q)$ at $q=x_n$, and where $P(q)$ corresponds to some smoothing of the empirical distribution of the world positions $\{x_1,x_2,\dots,x_N\}$. Equation~(\ref{ux}) may be regarded as an approximation of the quantum potential in Bohm's theory~\cite{Boh52_1,Boh52_2} (for which $P(q)=|\Psi(q)|^2$), and hence is expected to reproduce quantum evolution more and more closely as this approximation is improved. The  toy model potential in Equation~(\ref{utoy}) is the simplest such approximation~\cite{MIW14}. 
 
To go to higher order approximations, we tried two different methods.  The first we call the rational smoothing method and the second the equivariance method.  
In rational smoothing we take a systematic approach to approximate $P(q)$ and its derivative by ratios of polynomials, where the order of these polynomials determines the number of neighbours that each world directly interacts with. This approach is developed in the next subsection, followed by an example. It will be applied to the description of nodes in Section \ref{rrs}. rational smoothing is less computational resource-intensive compared to the equivariance Method. The latter is discussed in Appendix \ref{Equivariance}.

\subsection{Constructing Generalised Interworld Potentials}

 If \blk $P(q)$ is some smooth probability density  as above, \blk which approximates the distribution of world positions, $\{x_1,x_2,\dots,x_N\}$, then the cumulative distribution $C(x):=\int_{-\infty}^x dq \, P(q)$ must satisfy (at least approximately)
\beq \label{cx}
{\cal C}(x_n)=\int_{-\infty}^{x_n} dq\,P(q) = u_n:=\frac{n-\half}{N} 
\eeq

Thus, the area under $P(q)$ is divided into $N$ neighbourhoods, each containing one world and having area $1/N$.  It will be assumed for simplicity that $P(q)$ is nonzero almost everywhere, implying that ${\cal C}(x)$ is strictly monotonic and hence invertible.

To {systematically approximate the quantum force at $x_n$ to a given accuracy, we first define the inverse of the cumulative distribution by $y(u):={\cal C}^{-1}(u)$. It follows immediately from Equation~(\ref{cx}) that}
\beq \label{xn}
x_n = y(u_n) .
\eeq

Further,  differentiating  $u={\cal C}(y)$  with respect to  $y$ \blk and using $d/dy\equiv (1/y')(d/du)$, gives 
$P(y)=1/{y'}$, $P'(y)=-y''/(y')^3$, and hence that  
\beq \label{ratio}
\frac{P'(y)}{P(y)} = -\frac{y''}{(y')^2}.
\eeq  

Our aim is now to approximate this last expression by a function of the world positions, and then substitute this approximation into the right hand side of Equation~(\ref{ux}) to obtain a corresponding form for the interworld potential $U(X)$. 

In particular, expanding $y(u_{n+c})$ in a Taylor series about $y(u_n)$ gives the approximation

	\begin{equation} \label{taylor} 
	x_{n+c} -x_n = y(u_{n+c})-y(u_n) \approx \sum_{l=1}^L \frac{1}{l!}\left(\frac{c}{N}\right)^l y^{(l)}_n 
	\end{equation}
	\blk
to accuracy $O(1/N^{L})$, where $y^{(l)}_n$ denotes the $l$th derivative of $y(u)$ at $u=u_n$.   The first $L$ derivatives of $y$  can therefore be approximated to this accuracy by choosing a set of coefficients $\{ \alpha_{cl}\}$ such that 
	\begin{equation} \label{alpha}
	\sum_{c} \alpha_{cl}\,c^{l'} = l!\,\delta_{ll'},~~~~~l,l'=1,2,\dots,L.
	\end{equation}
	
It is shown how to construct suitable $\{ \alpha_{cl}\}$ below.   
Equations~(\ref{taylor})  and (\ref{alpha}) immediately yield the~approximations
	\begin{equation} \label{yderiv}
	\frac{y^{(l)}_n}{N^l} \approx \sum_c \alpha_{cl}\,(x_{n+c}-x_n)  ,
	\end{equation}
accurate to $O(1/N^{L})$, for $l=1,2,\dots,L$. Finally, substitution into Equation~(\ref{ratio}) and hence into Equation~(\ref{ux}) gives the corresponding interworld potential,
\vspace{6pt}
\begin{equation} \label{rational}
	U_{\alpha,L}(X) = \sum_{n=1}^N U_{\alpha,L,n}(X) = \frac{\hbar^2}{8m} \sum_{n=1}^N  \left\{ \frac{\sum_c \alpha_{c2}(x_{n+c}-x_n)}{\left[ \sum_c \alpha_{c1}(x_{n+c}-x_n)\right]^2}\right\}^2 .
	\end{equation}

Note that this interworld potential is translation invariant, and scales as $1/\lambda^2$ under $x_n\rightarrow\lambda x_n$.  It follows that analogues of Ehrenfest's theorem and wavepacket spreading hold for all such potentials~\cite{MIW14}.

An advantage of Hamiltonian based methods is that we can use a symplectic numerical integrator to make sure that the total energy is conserved \cite{canonical}. The quantum force for the mth world-particle can be calculated from Equations~(\ref{hameq})~and~(\ref{rational}) as: 
\begin{equation}
f_m=-\frac{\partial U_{\alpha,L}(X)}{\partial x_m}=  -\sum_{n=1}^{N} \frac{\partial U_{\alpha,L,n}(X)}{\partial x_m} .
\label{f_n_1}
\end{equation}

\subsection{Examples}
	
	To obtain an explicit example of the interworld potential in  Equation~(\ref{rational}), let $M$ denote an $L\times C$ matrix with coefficients $M_{lc}=c^l$, and $A$
	denote the $C\times L$ matrix with coefficients $\alpha_{cl}$, where $c$ ranges
	over some set of $C$ integers. 
	Equation~(\ref{alpha}) can then be written in the matrix form
	\beq 
	MA = \Delta:= {\rm diag}[1!,2!,\dots,L!] . 
	\eeq
	
	The existence of a solution requires $C\geq L$.  Further, since the values of $\alpha_{c1}$ and $\alpha_{c2}$ are required in Equation~(\ref{rational}), one must have $L\geq 2$.  The corresponding solution is then $A= M^{-1} \Delta$, where $M^{-1}$ denotes the inverse of $M$ for $C=L$, and a pseudo-inverse of $M$ for $C>L$.
	
	It follows that the simplest interworld potential constructed in this way corresponds to $C=L=2$.  Labelling the two values of $c$ by $c=\pm 1$ corresponds, via Equation~(\ref{yderiv}), to approximating the derivatives of $y(x_n)$ via the values of the nearest 
	neighbours $x_{n\pm1}$. 
	Ordering the values of $c$ as $-1, 1$, the corresponding coefficients $\{\alpha_{cl}\}$ in Equation~(\ref{rational}) are then given by
	\beq 
	A =M^{-1}\Delta =\left( \begin{array}{rr}
	-1  & 1  \\
	1 & 1 
	\end{array} \right)^{-1}
	\left( \begin{array}{cc}
	1  & 0  \\
	0 & 2 
	\end{array} \right)
	= \left(
	\begin{array}{rr}
	 -\xfrac12 & 1 \\
	\xfrac12 & 1
	\end{array}
	\right) .
	\eeq
	
 Surprisingly, this is not actually equivalent to the toy model of Reference~\cite{MIW14}, even though it involves the same number of neighbours in the potential. 

 The simplest higher-order interworld potential corresponds to $C=L=3$,  but odd values of $C$ necessarily introduce an unphysical left-right asymmetry, with each world being coupled to different numbers of neighbouring worlds on either side via $ U_{\alpha,L,n}(X)$.   To preserve symmetry we therefore next consider the case $C=L=4$. \blk Labelling the four values of $c$ by $c=\pm 1,\pm 2$, the derivatives of $y(x_n)$ in Equation~(\ref{yderiv}) can be calculated with values of the nearest and next nearest neighbours $x_{n\pm1}$ and $x_{n\pm2}$.  If we order the values of $c$ as $-2, -1, 1, 2$, the coefficients $\{\alpha_{cl}\}$ in Equation~(\ref{rational}) are then given by
 \vspace{6pt}
\begin{align} \nn
A =M^{-1}\Delta &=\left( \begin{array}{rrrr}
-2  & -1 & 1 & 2 \\
 4 & 1 & 1 & 4 \\ 
-8 & -1 & 1 & 8 \\ 
16 & 1 & 1 & 16 
\end{array} \right)^{-1}
\left( \begin{array}{cccc}
1  & 0 & 0 & 0 \\
0 & 2 & 0 & 0 \\ 
0 & 0 & 6 & 0 \\ 
0 & 0 & 0 & 24 
\end{array} \right)\\
&= \left(
\begin{array}{cccc}
 \xfrac{1}{12} & -\xfrac{1}{12} & -\xfrac{1}{2} & 1 \\
 -\xfrac{2}{3} & \xfrac{4}{3} & 1 & -4 \\
 \xfrac{2}{3} & \xfrac{4}{3} & -1 & -4 \\
 -\xfrac{1}{12} & -\xfrac{1}{12} & \xfrac{1}{2} & 1 \\
\end{array}
\right) . \label{simplestrational}
\end{align}

\section{Numerical Results} \label{sec6}

To test our higher-order methods against the original toy model, we apply them to the problem of finding the ground and the 
first excited state of a harmonic oscillator.  It is the latter, containing a node, where the toy model fails and 
it is necessary to use a higher-order interworld potential.  
 In this section we use dimensionless configuration coordinates
\begin{equation} \label{qndefD}
{X}_n := \sqrt{2m\omega/\hbar} \,\,x_n  
\end{equation}    
and dimensionless times   
\begin{equation} \label{tfD}
T := \omega t / 2 \pi ,
\end{equation} 
where $\omega$ is the harmonic oscillator frequency.

\subsection{Toy Model}

\begin{figure}[H]
\begin{center}
\includegraphics[width=1\columnwidth]{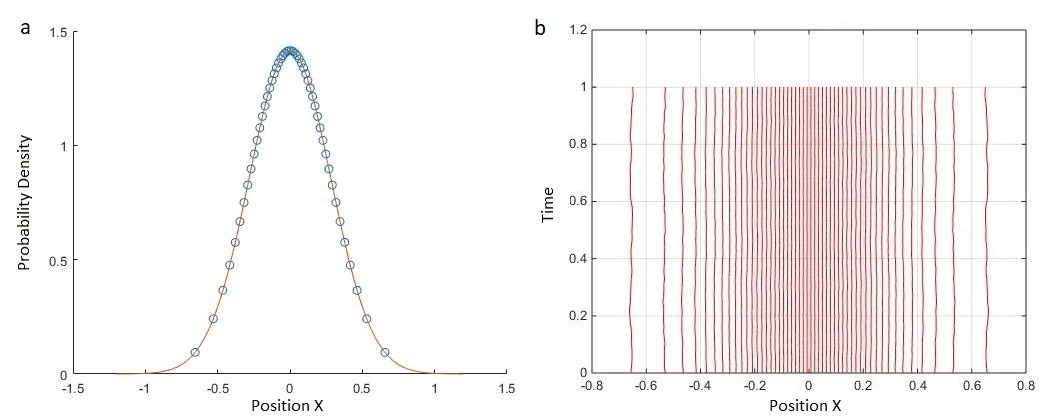}
\end{center}
\caption{(\textbf{a}) {50 world-particles}  (that is, 50 different worlds describing a single particle) are distributed based on the probability density of the ground state of a harmonic oscillator. (\textbf{b}) Trajectories of the 50~world-particles. As expected, the classical and quantum forces approximately cancel each other and { the world-particle}s stay approximately stationary. The slight oscillations near the boundary are due to approximation of the Bohmian potential by the toy-model potential.}
\label{G_S_prob_u3}
\end{figure}

Figure~\ref{G_S_prob_u3}b shows the result of applying the toy model in Equations~(\ref{hxp})~and~(\ref{utoy}) to the ground state of a harmonic oscillator, corresponding to $V(q) = \half m \omega^2 x^2$.  In this test we distributed 50 world-particles with $x_n$ determined by inverting Equation~(\ref{cx}) for the groundstate probability density $P_t^{(0)}(q)$ (Figure~\ref{G_S_prob_u3}a).  We then evolved these under the single-particle harmonic oscillator potential  $\half m \omega^2 x_n^2$ and the quantum  interworld potential. We see that as expected for a stationary quantum potential, the classical and quantum forces cancel each other and the world-particles stay stationary. The slight oscillations for the world-particles near the boundary is due to differences between the toy-model and Bohmian potentials in areas with high curvature and low sampling. These differences imply an exact stationary state for the MIW potential that has slightly different values for the $x_n$ for any finite value of $N$ \cite{MIW14}.

\begin{figure}[H]
\centering
\includegraphics[width=1\columnwidth]{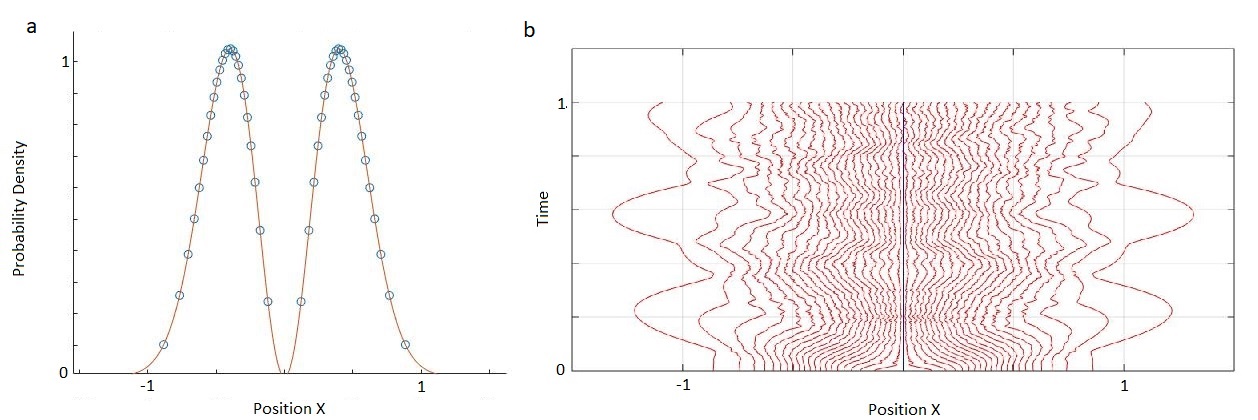}
\caption{(\textbf{a}) 40 world-particles are distributed based on the probability density of the first excited state of a harmonic oscillator. (\textbf{a}) Trajectories for the first excited state of a harmonic oscillator using the potential in Equation~(\ref{utoy}). The world-particles do not stay stationary. Particularly those near the node move towards the middle and fill the gap.}
\label{prob_1st_traj}
\end{figure}

However, if we distribute world-particles based on the probability density of the first excited state of a harmonic oscillator (Figure~\ref{prob_1st_traj}a) and apply the same model, the world-particles will not stay stationary as expected (Figure~\ref{prob_1st_traj}b).  We also implemented the nearest-neighbour interworld potential defined by Equation~(\ref{simplestrational}), and found very similar behaviour to the original toy model. That is, it fails to support stationary configurations corresponding to excited energy eigenstates. \blk 

\begin{figure}[H]
\centering
\includegraphics[width=1\columnwidth]{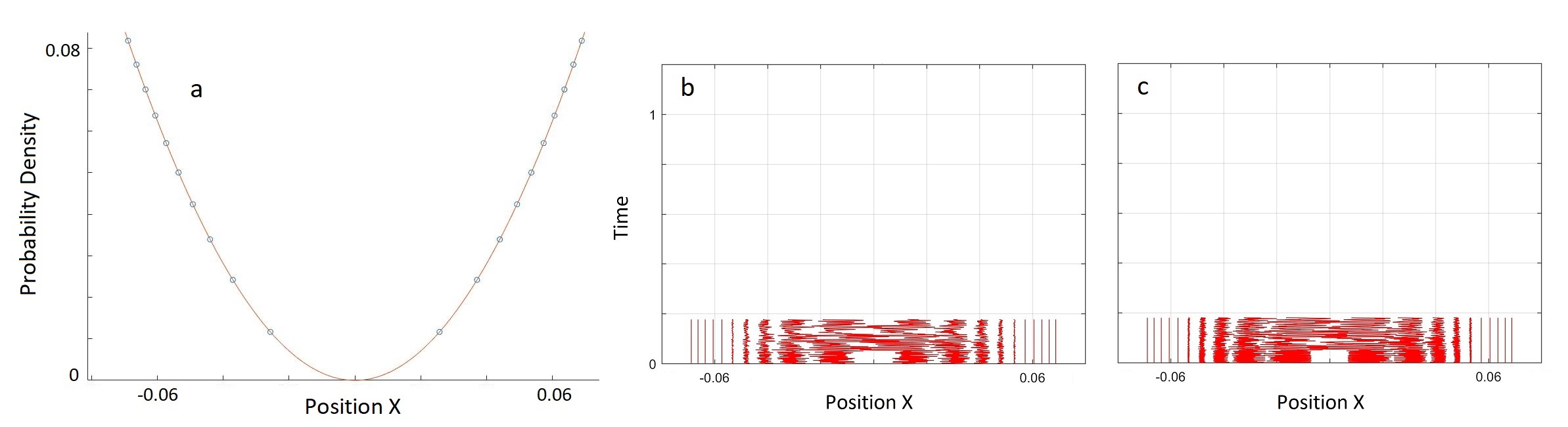}
\caption{(\textbf{a}) {5000 world-particles} are distributed based on the probability density of the first excited state of a harmonic oscillator. Only 20 world-particles around the node in the middle are shown. (\textbf{b})~Trajectories of the world-particles for the initial distribution in (\textbf{a}). To apply the correct boundary condition, we kept five world-particles, near each boundary, fixed and simulated the dynamics of the remaining 10 world-particles in the middle. time-steps are $10^{-8}$.  Since the nearest neighbours of the node do not stay near the starting point and move to the middle of the node after approximately 0.1 of a period, we did not continue the simulation for a full period. (\textbf{c}) The same test as (\textbf{b}) with time-steps of $10^{-9}$.  Thus, the failure of the simulation is not an artefact of large time-steps.}
\label{Probability_1st_5k_mid_traj}
\end{figure}

The unsuccessful simulation results in Figure~\ref{Probability_1st_5k_mid_traj} show that the problem of the nodes cannot be fixed by increasing the sampling. In the case of a node, because the probability density is zero, there is always a second order curve between the two world-particles adjacent to the node that can never be correctly estimated by toy model approximation. This leads to a poor approximation of the Bohmian force for the world-particles near the node and instead of staying stationary the world-particles move towards the gap and fill the gap.  

The {general problem with areas of low probability density,  and in particular in the region of the node of the quantum state at $x=0$ in Figure~\ref{prob_1st_traj}a, is that sampling is low and  nearest-neighbour approximations are not valid for calculating the quantum potential. If the probability is low but non-zero, theoretically, we can increase the number of worlds until we reach a good sampling in those areas. To test the simulation with higher sampling, we tried 5000 worlds. To reduce the simulation time, we only focused on the 20 world-particles around the node (Figure~\ref{Probability_1st_5k_mid_traj}a). To apply the correct boundary  condition for truncated area, we kept five world-particles near each boundary artificially fixed.  }
\unskip

\FloatBarrier

\subsection{Higher Order Potential}
\label{rrs}
To test a higher order approximation, we applied the rational smoothing model in Equation~(\ref{rational}) for the case $L=4$ (equivalent to a 5-world interaction), to the first excited state of a harmonic oscillator with 5000 worlds. This corresponds to the example in Equation~(\ref{simplestrational}).

\begin{figure}[H]
\centering
\includegraphics[width=0.75\columnwidth]{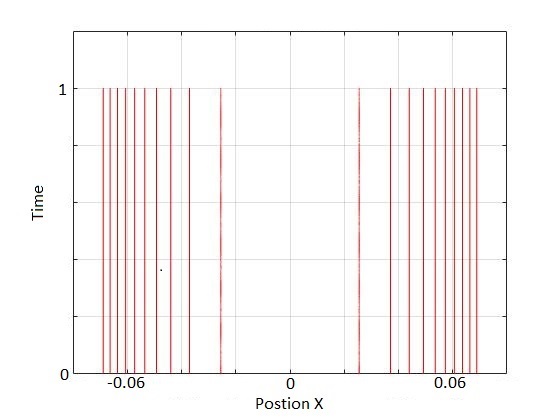}
\caption{{Simulation of only two worlds} adjacent to the node for the first excited state of a harmonic oscillator.  The initial positions were set by considering 5000 worlds in total to describe the excited state. To apply the boundary condition,  the rest of the world-particles were kept fixed.  For the evolution, rational smoothing with {$L = 4$} 
 (equivalent to 5-world approximation) is used. Time step is $10^{-9}$. The~two world-particles stay stationary as expected.}
\label{5k_1st_neighbours}
\end{figure}

In the first test we kept all the worlds artificially stationary except for the two worlds around the node. Figure~\ref{5k_1st_neighbours} shows that these two middle worlds stay stationary which means that the quantum potential for those is well approximated.
\vspace{-12pt}

\begin{figure}[H]
\centering
\includegraphics[width=0.75\columnwidth]{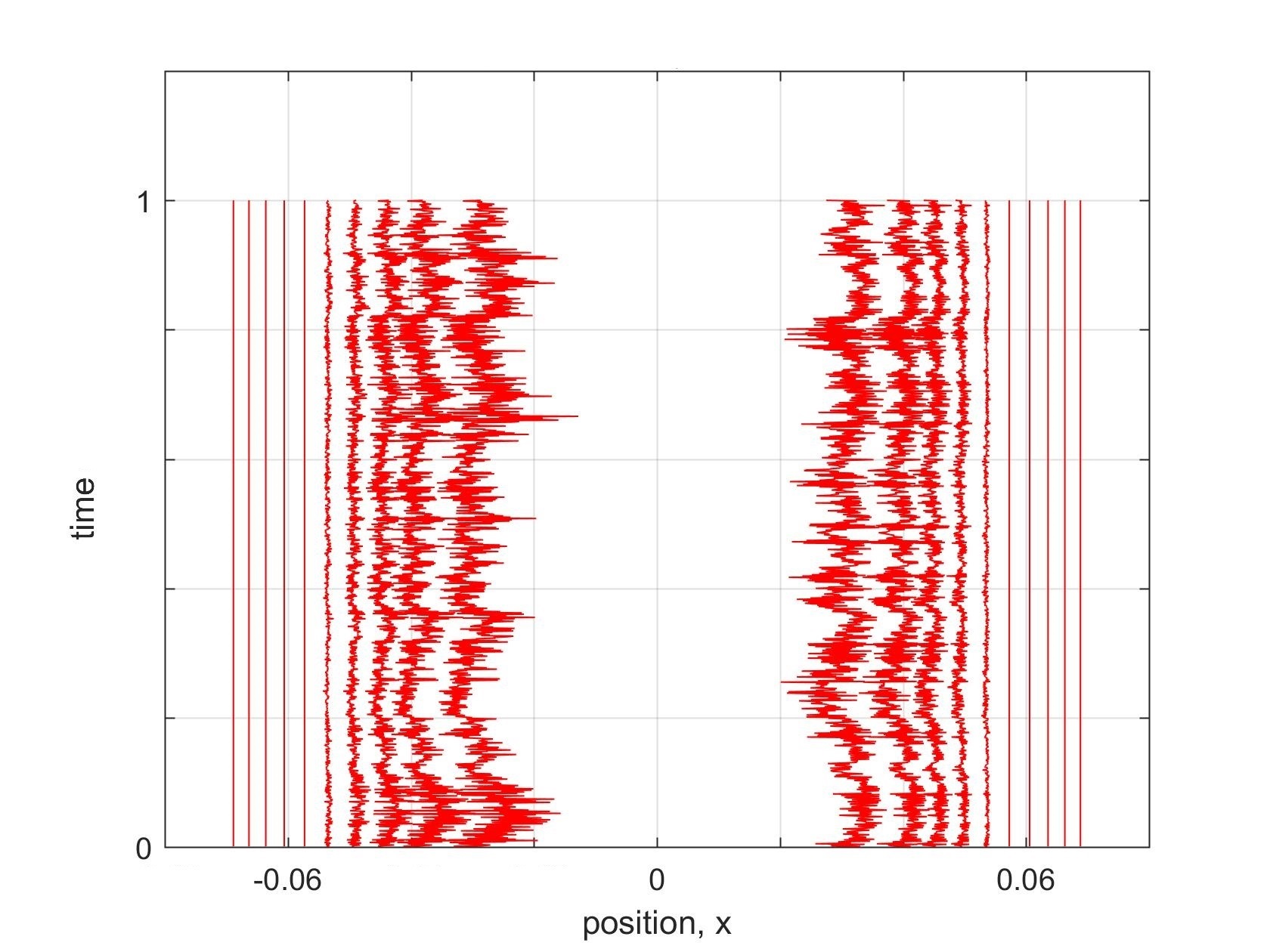}
\caption{{Simulation of 10 worlds}  (rather than two, as in Figure~\ref{5k_1st_neighbours}), five on either side of the node for the first excited state of a harmonic oscillator.  Other details are as in Figure~\ref{5k_1st_neighbours}.}
\label{5k_5_neighbours}
\end{figure}

In the second test we tried the same scenario but this time simulated 10 worlds around the node and kept the rest stationary. Figure~\ref{5k_5_neighbours} shows that the simulation was successful in the sense that world-particles stayed close to the starting position and did not move towards the gap i.e., the low density region, corresponding to a node of the associated excited state, is preserved by the interaction, and the evolution is approximately stationary. The obvious difference compared to Figure~\ref{5k_1st_neighbours} is the oscillations around the starting point.  The oscillations appear chaotic, as might be expected for a system of nonlinearly coupled harmonic oscillators. \blk
These oscillations  might also \blk be due to the time step being too long. We were not able to test this conjecture because the test in Figure~\ref{5k_5_neighbours} took a few days on our desktop computer and, due to time constraints, we could not run it with smaller time steps.  

\begin{figure}[H]
\centering
\includegraphics[width=0.75\columnwidth]{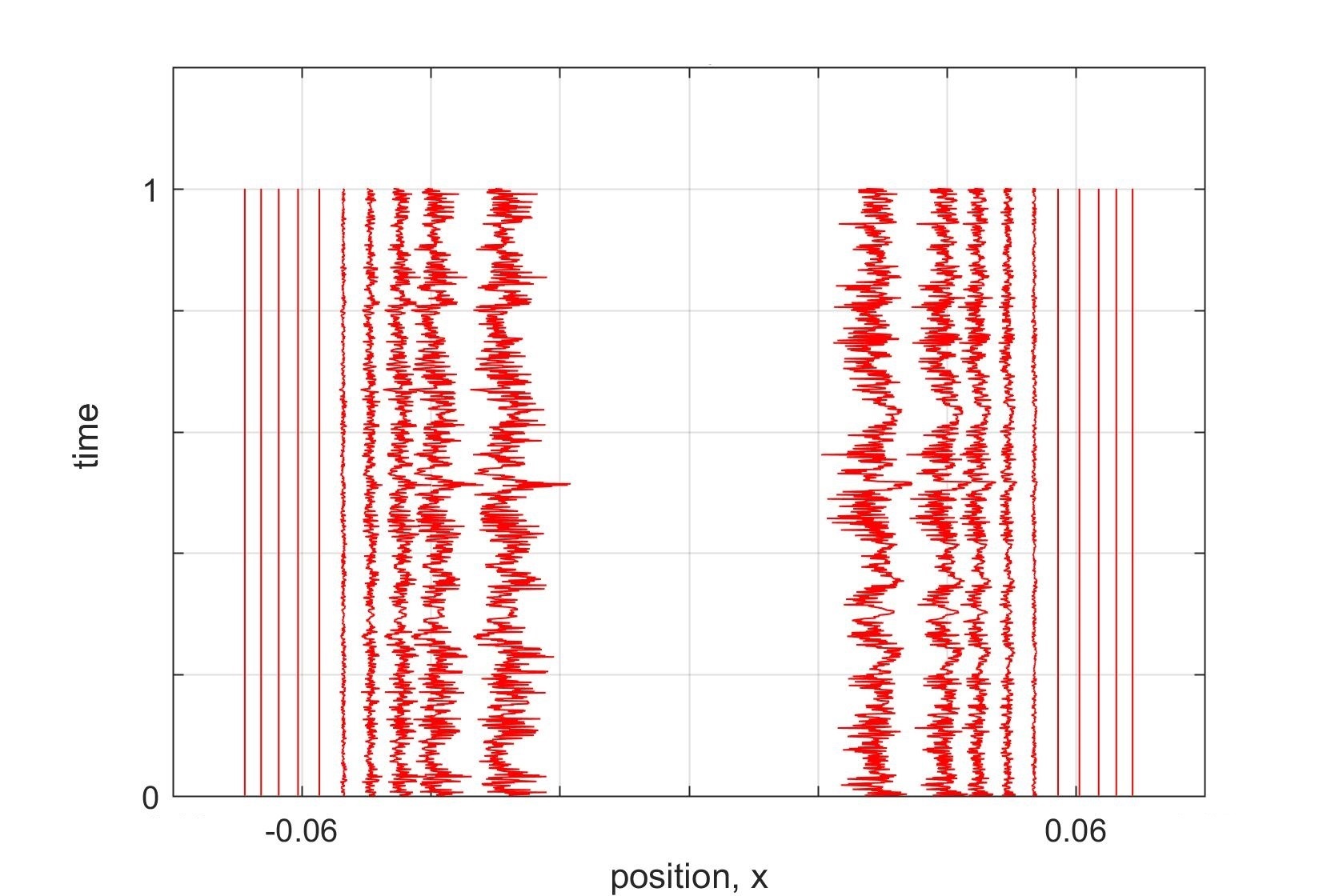}
\caption{ Simulation of the first excited state of a harmonic oscillator with rational smoothing, $L = 6$, (7-world approximation). Other details are as in Figure~\ref{5k_5_neighbours}.  The oscillations are much smaller than those in the $L=4$ {approximation in Figure}~\ref{5k_5_neighbours}. \blk}
\label{5k_5_neighbours_l7}
\end{figure}

We repeated the same test with 7-worlds approximation ($L=6$). Figure~\ref{5k_5_neighbours_l7} shows that this higher order approximation decreases oscillation compared to the results in Figure~\ref{5k_5_neighbours}. Hence, the simulation of a wave function with nodes, which failed for the toy model (Figures~\ref{prob_1st_traj} and \ref{Probability_1st_5k_mid_traj}), is seen to become more and more accurately modelled in the MIW approach as the number of directly interacting worlds is increased (Figures~\ref{5k_5_neighbours}~and~\ref{5k_5_neighbours_l7}).

Similar convergence might also be possible 
with the equivariance Method presented in Appendix~\ref{Equivariance}. However, the results of the simplest test, with a potential involving five interacting worlds, were not as positive as those of the rational smoothing explored in this section. They  are also more numerically intensive, so we did not pursue it further.  
\vspace{-18pt}

\FloatBarrier 
\section{Conclusions}\label{sec7}
 
Bell's theorem of 1964 showed that any deterministic interpretation of quantum mechanics must be nonlocal (Section~\ref{sec2}). 
Bohm's theory of 1952 is the example 
{\em par excellence} of a nonlocal, deterministic theory (Section~\ref{sec3}). The theoretical approach we introduced in 2014---Many Interacting Worlds---is also deterministic and, if it is to succeed in replicating Bohmian mechanics (and thus all quantum phenomena), 
must be nonlocal (Section~\ref{sec4}). 
It is conceptually clear how an interworld potential can lead to nonlocality in Bell's 1964 sense, and, perhaps surprisingly,
this could in principle be simulated with a universe of only two particles, in 1D (one spatial dimension), if an externally controlled 
 spatially localized \blk potential is allowed, as discussed in Section~\ref{sec4}. 

Unfortunately, this proof-of-principle simulation of Bell-nonlocality cannot be done using  
the toy model for an interworld potential introduced in Reference~\cite{MIW14}. The reason is that it cannot deal with wavefunctions having nodes, 
and nodes are certainly necessary for modelling the entangled states and measurements necessary for violating a Bell inequality~\cite{Bel86}. 
When one prepares a distribution of worlds, in one dimension, corresponding to a stable excited quantum state, the dynamics of the toy model causes the gap between worlds where the node should collapse, and no stationary configuration is reached.  Dealing with nodes is a problem in many quantum simulation methods based on {Bohmian mechanics}~\cite{node_wyatt}.  
Nodes should not be a problem for interpretations involving a continuum of worlds~\cite{Holland05,Poirier10,ParPoi12,SchPoi12}, as they are formulated to be exactly equivalent to quantum mechanics. However, as remarked in Section~\ref{sec3}, our view is that these interpretations do not solve the conceptual problems of the Everettian many-worlds interpretation.
\blk 

Here we showed that  this problem of nodes in  our discrete \blk MIW approach may not be fundamental, but rather may be a 
 result of using a too simple form for the inter-world interaction potential.  By using a higher-order approximation to define our 
 interworld potential from Bohm's quantum potential, we were able to show that a gap in the configuration of worlds, corresponding to the node of the first excited harmonic oscillator energy eigenstate, can remain open for at least a full harmonic oscillator period, and perhaps indefinitely. The world configurations were not exactly stationary, but rather had high frequency irregular oscillations. However, by increasing the order of the approximation, {the size of the oscillations could be reduced.} 

Our simulations considered only the dynamics near the node, with more distant worlds artificially held fixed. Whether the node would remain stable if all worlds were allowed to evolve according to the MIW dynamics is an open question. In addition, our simulations were restricted to one particle in 1D. The MIW approach has been successfully used to simulate one particle in 2D (or, equivalently, two particles in 1D)~\cite{HHWD17, Sturniolo18} and 3D~\cite{Sturniolo18},  but only to find the ground state configuration. Combining~these research directions to be able to simulate stable excited states for two particles in 1D, and beyond, is a challenge for future work. Finally, to realise a simulation of the Bell experiment described in Section~\ref{sec4} would require simulating not just stationary nodes but also {\em dynamical} 
 nodes, that may appear and then disappear in an instant, which are notoriously difficult to deal with in Bohmian-inspired numerical approaches~\cite{node_wyatt}.  
Thus, much work remains to be done, but the positive results reported here are encouraging. \blk

\authorcontributions{ H.M.W. and M.J.W.H. 
conceived the project to demonstrate Bell-nonlocality in the MIW approach.  M.J.W.H. developed the  rational smoothing method.  M.G.  developed  the equivariance method and performed the simulations, with discussions involving all authors. H.M.W. drafted Sections \ref{sec:in}--\ref{sec4} and \ref{sec7}, M.J.W.H. drafted Section \ref{sec5}, M.G. drafted Section \ref{sec6}  and the Appendix.  All authors contributed to the final form of the paper.}
 
{\blk This research was funded by the Foundational Questions Institute grant number FQXi-RFP-1519.}
 
\acknowledgments{We thank Dirk-Andr\'e Deckert for his formative and continuing role in MIW theorising with us, and the hospitality of Ludwig Maximilian University in facilitating the collaboration. 
  H.M.W.\ and M.J.W.H.\ acknowledge Evan Gamble for valuable discussions on simulating excited states in the MIW approach, some years ago now. 
H.M.W.\ also thanks the Fetzer \blk Franklin \blk fund for support to 
attend the David Bohm Centennial Symposium 
at which some of the ideas in Section~\ref{sec2} solidified in dialogue with other invitees. }
 
 {\blk The authors declare no conflict of interest. The funding sponsors had no role in the design of the study; in the collection, analyses, or interpretation of data; in the writing of the manuscript, and in the decision to publish the results.}

\newpage

\abbreviations{The following abbreviations are used in this manuscript:\\

\noindent  
\begin{tabular}{@{}ll} 
MIW & Many Interacting Worlds\\
EPR & Einstein, Podolsky, and Rosen\\ 
1D & One (spatial) dimension\\
2D & Two (spatial) dimensions\\
3D & Three (spatial) dimensions 
\end{tabular}}
\blk

\appendix

\section{A Equivariance Method} \label{Equivariance}
Here we introduce an alternative higher order method to rational smoothing, motivated by the equivariance property of Bohmian mechanics \cite{DGZ92}. In particular, for a given configuration of worlds, $\{x_1, x_2, \dots ,x_n\}$, we construct a smooth polynomial probability density, $P_n(x)$, in the region of each world $x_n$, and use this to approximate the Bohmian potential.   The coefficients of the polynomial are determined by requiring that area under $P_n(x)$ between $x_n$ and $x_{n+1}$ is equal to the constant value $1/N$ for each world, analogously to Equation~(\ref{cx}).
The accuracy of this equal-probability or ``equivariance'' method will increase with the degree of the polynomial.  

Here we will illustrate the equivariance method for the case where $P_n(x)$ is third-order, corresponding to direct interactions between sets of five adjacent worlds. Thus,
\begin{equation}
P_n(x)=a_n+b_nx+c_nx^2+d_nx^3.
\label{prob_eq}
\end{equation}

The equivariance method then requires that 
\begin{equation}
\int_{x_n}^{x_{n+1}} P_n(x) dx = \text{ const.} = N^{-1}.
\end{equation}

To determine the coefficients of $P_n(x)$ in Equation~(\ref{prob_eq}), we use four equations below based on the positions of the five worlds, $x_{n-2}$, $x_{n-1}$, $x_{n}$, $x_{n+1}$ and $x_{n+2}$:  
\begin{equation}
\int_{x_{n-2}}^{x_{n-1}} P_n(x) dx = N^{-1}
\end{equation}
\begin{equation}
\int_{x_{n-1}}^{x_{n}} P_n(x) dx = N^{-1}
\end{equation}
\begin{equation}
\int_{x_{n}}^{x_{n+1}} P_n(x) dx = N^{-1}
\end{equation}
\begin{equation}
\int_{x_{n+1}}^{x_{n+2}} P_n(x) dx = N^{-1}
\end{equation}
 
Substituting $P_n(x)$ from Equation~(\ref{prob_eq}) and {evaluating the integrals we get}
\begin{equation}
\begin{array}{cc}
a_n \, (x_{n-1}-x_{n-2})  + b_n \, \frac{1}{2} (x^2_{n-1}-x^2_{n-2}) + c_n \, \frac{1}{3} (x^3_{n-1}-x^3_{n-2}) + d_n \, \frac{1}{4} (x^4_{n-1}-x^4_{n-2})= N^{-1} \\
a_n \, (x_{n}-x_{n-1})  
+ b_n \, \frac{1}{2} (x^2_{n}-x^2_{n-1}) 
+ c_n \, \frac{1}{3} (x^3_{n}-x^3_{n-1}) 
+ d_n \, \frac{1}{4} (x^4_{n}-x^4_{n-1})= N^{-1} \\
a_n \, (x_{n+1}-x_{n})  
+ b_n \, \frac{1}{2} (x^2_{n+1}-x^2_{n}) 
+ c_n \, \frac{1}{3} (x^3_{n+1}-x^3_{n}) 
+ d_n \, \frac{1}{4} (x^4_{n+1}-x^4_{n})= N^{-1} \\
a_n \, (x_{n+2}-x_{n+1})  
+ b_n \, \frac{1}{2} (x^2_{n+2}-x^2_{n+1}) 
+ c_n \, \frac{1}{3} (x^3_{n+2}-x^3_{n+1}) 
+ d_n \, \frac{1}{4} (x^4_{n+2}-x^4_{n+1})= N^{-1}.
\label{int_lin}
\end{array}
\end{equation}

{If we define matrix $K_n$ as}
\begin{align}
K_n &=\left( \begin{array}{rrrr}
(x_{n-1}-x_{n-2})  & \frac{1}{2} (x^2_{n-1}-x^2_{n-2}) & \frac{1}{3} (x^3_{n-1}-x^3_{n-2}) & \frac{1}{4} (x^4_{n-1}-x^4_{n-2})\\ 
(x_{n}-x_{n-1}) & \frac{1}{2} (x^2_{n}-x^2_{n-1})& \frac{1}{3} (x^3_{n}-x^3_{n-1}) & \frac{1}{4} (x^4_{n}-x^4_{n-1})\\
(x_{n+1}-x_{n}) & \frac{1}{2} (x^2_{n+1}-x^2_{n})  & \frac{1}{3} (x^3_{n+1}-x^3_{n}) 
& \frac{1}{4} (x^4_{n+1}-x^4_{n}) \\ 
(x_{n+2}-x_{n+1}) & \frac{1}{2} (x^2_{n+2}-x^2_{n+1}) & \frac{1}{3} (x^3_{n+2}-x^3_{n+1}) & \frac{1}{4} (x^4_{n+2}-x^4_{n+1})
\end{array} \right),
\end{align}
we can rewrite Equation~(\ref{int_lin}) as
\begin{align}
K_n \left( \begin{array}{rrrr} a_n\\ b_n\\ c_n\\ d_n\\  \end{array} \right) = \left( \begin{array}{rrrr} N^{-1}\\ N^{-1}\\ N^{-1}\\ N^{-1} \end{array} \right).
\end{align}

Therefore:
\begin{align}
\left( \begin{array}{rrrr} a_n\\ b_n\\ c_n\\ d_n\\  \end{array} \right) = K_n^{-1}\left( \begin{array}{rrrr} N^{-1}\\ N^{-1}\\ N^{-1}\\ N^{-1} \end{array} \right).
\label{abcd}
\end{align}

We used {Matlab to invert $K_n$ and find coefficients $a_n, b_n, c_n,$ and $d_n$. The resulting equations were too lengthy to include here. Substituting these coefficients into Equation~(\ref{prob_eq}), we can find the~probability.  }

To evaluate the quantum force, we use an interworld interaction potential of the form of Equation~(\ref{ux}) and $P_n$ as in Equation~(\ref{prob_eq}) (with $a_n$, $b_n$, $c_n$ and $d_n$ taken from (\ref{abcd})) to give 

\begin{equation}
U=\frac{\hbar^2}{8m} \sum_{n=1}^{N} \left(  \frac{b_n+2c_nx_n+3d_nx_n^2}{a_n+b_nx_n+c_nx_n^2+d_nx_n^3}  \right)^2. 
\end{equation}

This may be compared with the corresponding interaction potential $U_{\alpha,L}$ in Equation~(\ref{rational}) obtained via rational smoothing.  The quantum force on each particle is evaluated similarly to Equation~(\ref{f_n_1}). We~used Matlab to evaluate the analytical derivatives of these terms for the simulation. The~resulting equation is too lengthy to include here  ({\em c.} 360,000 characters for $N=5$). 

\begin{figure}[H]
\centering
\includegraphics[width=0.6\columnwidth]{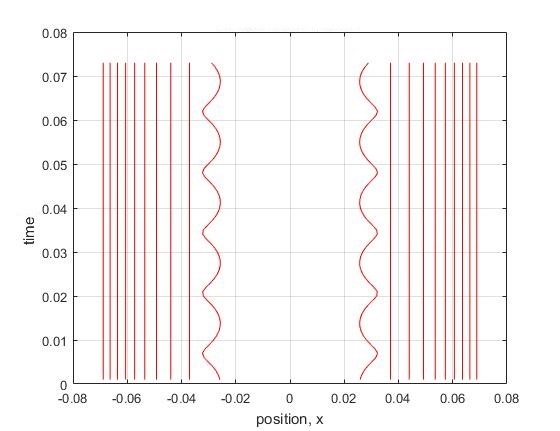}
\caption{{The first excited state of a harmonic oscillator} is simulated using 5-world approximation in equivariance Mmethod. Five-thousand worlds are used and only the two world-particles next to the node are simulated. The rest are kept stationary, similar to Figure~\ref{5k_1st_neighbours} for the rational smoothing case.}
\label{equi_5k_1st_neighbours}
\end{figure}

Figure~\ref{equi_5k_1st_neighbours} shows the result of applying the equivariance method to the first neighbours of the node in the first excited state of a harmonic oscillator. It shows that these world-particles do not move into the gap from their original position, but rather undergo some oscillatory motion. This is poorer behaviour than the corresponding rational smoothing simulation shown in Figure~\ref{5k_1st_neighbours}.  The simulations were also slower because of the complexity of the analytical form of the force law, mentioned above.  For these reasons we have not pursued this method further.  

\bibliographystyle{MDPI}
\bibliography{QMCrefsPLUS}

\end{document}